\documentclass[11pt]{article}
\usepackage{graphicx}
\usepackage{amsmath}
\usepackage[british]{babel}
\usepackage[bf]{caption}  
\usepackage[square]{natbib}  

\begin{document}

\begin{centering}
\parbox{\textwidth}{\huge \centering \bf Applying machine learning to improve simulations of a chaotic dynamical system using empirical error correction} \\[0.5cm]

\parbox{\textwidth}{\Large \centering Peter A. G. Watson} \\[0.5cm]
\parbox{0.75\textwidth}{\large \centering Atmospheric, Oceanic and Planetary Physics, University of Oxford, Oxford, UK.}

\end{centering}

\abstract{Dynamical weather and climate prediction models underpin many studies of the Earth system and hold the promise of being able to make robust projections of future climate change based on physical laws. However, simulations from these models still show many differences compared with observations. Machine learning has been applied to solve certain prediction problems with great success, and recently it's been proposed that this could replace the role of physically-derived dynamical weather and climate models to give better quality simulations. Here, instead, a framework using machine learning together with physically-derived models is tested, in which it is learnt how to correct the errors of the latter from timestep to timestep. 
This maintains the physical understanding built into the models, whilst allowing performance improvements, and also requires much simpler algorithms and less training data. This is tested in the context of simulating the chaotic Lorenz '96 system, and it is shown that the approach yields models that are stable and that give both improved skill in initialised predictions and better long-term climate statistics. Improvements in long-term statistics are smaller than for single time-step tendencies, however, indicating that it would be valuable to develop methods that target improvements on longer time scales. Future strategies for the development of this approach and possible applications to making progress on important scientific problems are discussed.}

\section{Introduction} \label{sec:intro}
Numerical weather prediction and climate models attempt to predict and simulate components of the Earth system, including the atmosphere and perhaps also the oceans, land surface and biosphere. Whilst the fundamental physical equations governing the system are known, they cannot be solved accurately with available computational resources. Instead, approximations are made in the models' equations, and this gives rise to errors in their output. Methods to reduce these errors are highly valuable for giving better warning of major meteorological and climatic events. 

Recently, great advances in machine learning have taken place, for example in the domains of image recognition and game-playing \citep[e.g.][]{He2016,Silver2017}. The algorithms developed have been found to excel at certain problems that involve predicting an unknown value given values of predictor variables (for example, predicting what objects a photograph contains given its pixel values)---this is similar to the problem of predicting future behaviour of the Earth system given knowledge of its past and present state, and so there has been high interest in applying machine learning to improve such predictions. This has included predicting future weather events directly from observations and post-processing dynamical models' output \citep[e.g.][]{Krasnopolsky2012,McGovern2017,Rasp2018b,Scher2018a}. 

Another emerging application is applying machine learning to improve components of the dynamical Earth system models themselves, particularly the parameterisations of unresolved small-scale processes such as radiative interactions and cloud processes. This could allow larger improvements in prediction skill than is achieved by post-processing models' output, by better representing the physical interactions between variables. 

Previous work has primarily used artificial neural networks (ANNs), which are functions constructed from ``neurons''. Neurons are simply functions that linearly combine their inputs and then apply a given (generally non-linear) transformation to produce the output value. ANNs pass input data into neurons, whose output may then be used as inputs to more neurons, and so on until a final output value (or vector of values) is produced. ANNs can relatively efficiently encode complex functional relationships. Indeed, an ANN with neurons arranged in layers, with the outputs of neurons in one layer being inputs to neurons in the next layer, can represent any real continuous function to an arbitrarily small level of accuracy, given a sufficient number of neurons \citep{Cybenko1989}. \citet{Nielsen2015} and \citet{Goodfellow2016} provide general introductions to the theory and applications of ANNs.

One promising approach has been to use algorithms such as ANNs to reproduce the behaviour of atmospheric parameterisation schemes at a reduced computational cost. For example, \citet{Chevallier1998,Chevallier2000b} found that ANNs could cheaply reproduce the behaviour of the radiative transfer scheme in the European Centre for Medium-Range Weather Forecasts model, although \citet{Morcrette2008} note that this approach did not work sufficiently well after the model's vertical resolution was increased. \citet{Krasnopolsky2005,Krasnopolsky2010a} also found that an ANN could be used to cheaply reproduce the output of the radiation scheme in the National Center for Atmospheric Research (NCAR) Community Atmosphere Model. More recent work has focussed on replacing atmospheric models' convection schemes with ANNs, in order not just to reduce the cost of presently used schemes but to allow more expensive, higher quality schemes to be used, such as superparameterisation \citep{Gentine2018,Rasp2018a} or emulations of convection-resolving models \citep{Brenowitz2018}. \citet{OGorman2018} also showed that a model's convection scheme could be replaced by a random forest algorithm, and the model could run stably and reasonably reproduce precipitation extremes. 

Machine learning also holds promise of being able to reduce errors in dynamical models' predictions. \citet{Schneider2017} describe how better values of parameters of models could be learnt by algorithms being fed data from observations and high-resolution models. \citet{Dueben2018} examine whether ANNs could be trained to simulate atmospheric dynamics and provide prediction skill exceeding that of existing models, using a time-stepping scheme where ANNs predict the tendency of the system in a similar approach to that used in existing dynamical models, and they conclude that it is possible. However, their simulations became unstable after about two weeks. Additionally, \citet{Bolton2019} showed that a convolutional ANN could skilfully predict small-scale momentum forcing in a quasi-geostrophic ocean model given spatially smoothed output from a simulation, although this was not implemented into a freely-evolving model. Significant advances have also been made in learning symbolic equations and invariant quantities from data \citep{Schmidt2009,Rudy2017,Wu2018,Zhang2018}, which could be applied in Earth system modelling \citep{Gaitan2016}.

The above-mentioned property of ANNs that they can represent any continuous function means that, in principle, given sufficient data of high enough quality to learn from and adequate computational resources for training, an ANN's representation of the equations of motion of the Earth system could reach the maximum skill possible for given inputs. So it seems that ANNs could potentially learn to reduce systematic model errors without the need for human ingenuity, greatly speeding up model development and helping us to address challenges like predicting extreme weather and the impacts of climate change. 

However, the use of ANNs as discussed in \citet{Schneider2017} and \citet{Dueben2018} has been presented as being in competition with improving the conventional physically-derived aspects of Earth system models. \citet{Schneider2017} argue that improving physically-derived parameterisation schemes is preferable to using ANNs because they will obey conservation laws and symmetries. \citet{Dueben2018} ask whether models based entirely on ANNs can compete with physically-derived models.

One purpose of the work presented here is to explore whether it is actually possible to use such algorithms to complement physically-derived model components, thereby preserving the benefits of using the latter, such as having better physical interpretability of the model behaviour and better trust that the model will perform reasonably well in an unseen physical situation. \citet{Karpatne2017a} and \citet{Reichstein2019} provide overviews of the ways in which statistical algorithms can be combined with physical modelling and the associated challenges. The specific proposal tested here is to use algorithms to perform empirical error-correction in dynamical models. Rather than predict the whole tendency of a system, as in the models considered by \citet{Dueben2018}, the algorithms would predict the difference between the measured and the observed tendencies. Then the total tendency would be $\mathcal{M}(x)+\epsilon(x)$, where $x$ is the system state at, and potentially before, the start of the time step, $\mathcal{M}(x)$ is the tendency predicted by the physically-derived model and $\epsilon(x)$ is the correction output by the algorithm. If $\mathcal{M}(x)$ is close to the optimum tendency, $\epsilon(x)$ should be small, and so concerns about $\epsilon(x)$ not obeying conservation laws and symmetries are consequently less important than in the case where whole model components are replaced by algorithms (note, though, that it may also be possible to constrain $\epsilon(x)$ to obey these physical principles more strictly \citep[e.g.][]{Jia2018}). $\epsilon(x)$ should only have a large effect on the simulations when the prediction by the physically-derived model is poor, in which case the value of improving the total simulated tendency is larger compared to concerns about whether physical principles are strictly abided by. The physically-derived model maintains a key role, and it is desirable to continue improving it to strengthen the link between the simulation results and our physical understanding. This study focuses on the use of ANNs as model error-correctors, but other algorithms could also be applied in a similar way.

A further advantage of using algorithms to correct models' errors rather than replace physically-derived models entirely is that it greatly simplifies the process of incorporating ANNs into dynamical models. \citet{Dueben2018} detail the numerous challenges in replacing physically-derived models with ANNs (or other algorithms), such as obtaining the required data for training a full-complexity model and learning to use algorithms with the required complexity. A lot of development effort would be required before a model with better performance than current models would be produced. By contrast, development of error-correcting algorithms can start just by improving a small number of outputs as much as possible given a small number of inputs, which is achievable with a smaller research programme, and progress can build from there. A disadvantage of this approach is that the computational cost of the models cannot easily be reduced this way if the resolution and parameterisation schemes are kept the same---the focus is on improving the simulation quality. However, it may turn out to be more cost-effective than using more expensive parameterisation schemes or increasing the model resolution, and it could reduce costs if it allows the same or greater skill to be obtained using cheaper physically-derived parameterisations. It would also be very informative about the problems that would need to be overcome to get dynamical models based entirely on algorithms like ANNs to perform well.

An empirical error-correcting approach using an ANN was applied by \citet{Forssell1997} to predict the water level in a tank. Previous environmental applications include the prediction of groundwater flow by \citet{Xu2015} and the prediction of lake temperatures by \citet{Karpatne2017b} and \citet{Jia2018}. These systems exhibit variability that is strongly influenced by external drivers, whilst Earth's atmosphere and oceans have a large component of unforced variability due to chaotic dynamics. A demonstration that using an algorithm to correct model errors could improve short-range forecasts of simple chaotic systems was given by \citep{Pathak2018}, who used reservoir computing, but it is unclear if this would also produce stable simulations with improved long-term statistics. \citet{Cooper2015} applied this approach to improve a chaotic shallow water model, but did so using a linear system of equations, which seems unlikely to be able to represent errors in complex dynamical systems as well as a more flexible algorithm such as an ANN in general---for example, atmospheric convection is a highly nonlinear process \citep{Arakawa2004}, and errors in its representation seem unlikely to be captured well by linear equations. Their work also focussed on improving long-term statistics of the simulation, and it is not clear whether there was a simultaneous improvement in short-range forecast skill that would indicate that the dynamics were being better represented. 

It is a key problem in Earth system prediction to improve our models in such a way that they are stable and give both improved skill in initialised predictions and better long-term climate statistics. In the remainder of this paper, it is tested whether this can be achieved by using an error-correcting ANN to better simulate a chaotic system, namely the Lorenz '96 dynamical system \citep{Lorenz1996} (sometimes also referred to as the Lorenz '95 system \citep[e.g. by][]{Dueben2018}). This system, or variants of it, has been used in many previous studies to test concepts for how to improve dynamical Earth system models \citep[e.g.][]{Wilks2005,Arnold2013,Schneider2017,Dueben2018}. The results are informative about the potential for machine learning approaches to improve skill at simulating dynamical systems such as the Earth's climate, albeit in a much simpler setting. 

\section{Experiments with the Lorenz '96 system} \label{sec:L96exp}

\subsection{The Lorenz '96 equations and coarse-resolution models}
The Lorenz '96 dynamical equations describe the evolution of variables arranged in a ring, intended to be analogous to a latitude circle. The variables are divided into two types: slowly-varying $X_{k}$ and quickly-varying $Y_{j,k}$, defined for $k=1,\ldots,K$, and $j=0,\ldots,J+2$. \citet{Lorenz1996} suggested that the $Y_{j,k}$ be considered analogous to a convective-scale quantity in the real atmosphere and $X_{k}$ analogous to an environmental variable that favours convective activity. Here one of the systems used by \citet{Arnold2013} is simulated as the ``Truth'' system---there is no particular strong reason to choose this system over other variants, but its prior use in the parameterisation development work of \citet{Arnold2013} makes it seem like a good choice for exploring how the design of models can be further advanced. This has $K=8$ and $J=32$, in which:
\begin{equation} \label{eqn:L96}
\begin{split}
\frac{\mathrm{d}X_{k}}{\mathrm{d}t} = -X_{k-1}(X_{k-2} - X_{k+1}) - X_{k} +F - (hc/b) \sum_{j=1}^{J} Y_{j,k}, \\
\frac{\mathrm{d}Y_{j,k}}{\mathrm{d}t} = -cbY_{j+1,k}(Y_{j+2,k} - Y_{j-1,k}) - cY_{j,k} + (hc/b)X_{k},
\end{split}
\end{equation}
with cyclic boundary conditions $X_{k}=X_{k+K}$ and $Y_{j,k}=Y_{j,k+K}$ and parameter values $h=1$, $F=20$, $b=10$ and $c=4$. The $Y$ variables are connected in a ring, such that $Y_{0,k}=Y_{J,k-1}$, $Y_{J+1,k}=Y_{1,k+1}$ and $Y_{J+2,k}=Y_{2,k+1}$, so there are $J$ unique $Y_{j,k}$ variables associated with each $X_{k}$ variable. The time units are arbitrary and denoted as model time units (MTUs). These equations were integrated in time with a time step of 0.001MTU using a fourth-order Runge-Kutta time-stepping scheme.

A ``training'' simulation of this system of length 3000MTUs (not including 10MTUs discarded as `spin up' at the beginning) was produced to provide a sample of ``true'' statistics to use in constructing coarse-resolution models below. The simulation was then extended for 10MTUs so that memory of the training dataset was effectively lost, and then a further 3000MTUs of data was generated to use as a validation dataset, which is used only to evaluate and not to develop the coarse-resolution models below. (Note that evaluation against a separate ``test'' dataset is not done because the aim is not to select a single best-performing ANN structure and estimate the model's true skill. This means that an ANN that is selected on the basis of having an especially good performance on the validation dataset would not be expected to perform as well relative to other ANNs on separate test data, since sampling variability would be expected to have contributed to its diagnosed skill in the validation data. It is shown below that performance improvements on the validation data are obtained for a wide range of ANN structures, and so the conclusion that ANNs can improve prediction skill for this system is robust to sampling variability.)

\subsubsection{Coarse-resolution model}
Suppose that a much computationally cheaper model of equations~\ref{eqn:L96} is desired for making short-term forecasts and simulating the long-run statistics of this system. Following \citet{Wilks2005} and \citet{Arnold2013}, inspection of equations~\ref{eqn:L96} suggests that it may be reasonable to forego simulating the $Y$ variables explicitly and parameterise their effect on the $X$ variables with a function $U(X)$, analogous to how the effect of unresolvable physical processes on resolved scales is parameterised in Earth system models. This yields the coarse-resolution system
\begin{equation} \label{eqn:L96_coarse}
\frac{\mathrm{d}X^{*}_{k}}{\mathrm{d}t} = -X^{*}_{k-1}(X^{*}_{k-2} - X^{*}_{k+1}) - X^{*}_{k} +F - U(X^{*}_{k})
\end{equation}
with $X_{k}=X_{k+K}$. The time step is also increased to 0.005MTU, so that the system has a coarsened time resolution as well, analogous to how Earth system models cannot resolve real Earth processes that happen on very fast time scales. 

The function $U(X^{*}_{k})$ is derived using the same method as \citet{Arnold2013}, using essentially a coarse-graining approach. It is defined as a cubic function, 
\[ U(X)=\sum_{n=0}^{3} a_{n}X^{n}, \] 
and its parameters are chosen using tendencies of the $X$ variables over intervals of length 0.005MTU, derived from the run of the truth system sampled every 0.005MTU. Its parameters were fit to minimise the root mean square error of predictions of these tendencies made using equation~\ref{eqn:L96_coarse}, taking values $a_{0}=-0.207$, $a_{1}=0.577$, $a_{2}=-0.00553$ and $a_{3}=-0.000220$. This is a statistical procedure, but note that here $U(X)$ is not considered part of the machine learning algorithm used to correct the model errors. Following \citet{Wilks2005} and \citet{Arnold2013}, $U(X)$ is thought of as being analogous to parameterisations of unresolved processes in an Earth system model---note that development of physically-derived Earth system model parameterisations can also involve fitting parameters to data \citep[e.g.][]{Hourdin2016}. Including $U(X)$ in equation~\ref{eqn:L96_coarse} helps to test whether complex algorithms such as ANNs are able to give skill improvements after much of the possible progress has been made with physical reasoning and simpler statistical methods, using the deterministic model of \citet{Arnold2013} as a benchmark. It would also be possible to do a similar study with $U(X)=0$, which would probably increase the potential improvement made by using error-correcting algorithms as they learnt to represent the improvements made by including $U(X)$.

Hereafter the model given by equation~\ref{eqn:L96_coarse} is referred to as ``No-ANN''.

\subsubsection{Models with ANNs} \label{sec:L96exp:ANNmods}
To produce coarse-resolution models with error-correcting ANNs, ANNs with a multilayer perceptron architecture \citep{Nielsen2015,Goodfellow2016} were trained to predict the difference between the true system tendency and that predicted by the coarse-resolution model for one $X$ variable at a time:
\[ \epsilon_{k} = \frac{\mathrm{d}X_{k}}{\mathrm{d}t} - \frac{\mathrm{d}X^{*}_{k}}{\mathrm{d}t}. \] The inputs to the ANNs are $X$ variables up to two points away from the location where the prediction is being made, so that five $X$ values in total are used as input. This is one more than used by the No-ANN model, and takes advantage of the ability of ANNs to use inputs that are difficult to know how to include by physical reasoning---this may be helpful in Earth system modelling to account for subgrid phenomena that propagate between grid boxes but are difficult to include in parameterisation schemes, such as horizontally-propagating gravity waves \citep{Alexander2010}. 
Results for ANNs using the same inputs as the No-ANN model are discussed at the end of section~\ref{sec:res} to quantify the impact made by using $X_{k+2}$ as an additional input---it does not qualitatively affect the findings. Not all of the $X$ variables were used as input in order that the ANNs are only using information from nearby grid points. This is desirable in Earth system models so they can be run much more quickly in parallel computing environments \citep{Dueben2018}. Also, since on Earth phenomena at one location could not be meaningfully influenced by phenomena on the other side of the world within one model time step, this structure constrains the ANNs to be more faithful to the true equations, so they are more likely to work well in novel situations. The results presented here are not expected to depend qualitatively on the number of $X$ variables used, and they were found to not be sensitive to using $X$ variables up to three points away instead. 

The $X$ variables are transformed by subtracting their mean and dividing by their standard deviation before being used as ANN inputs, since this tends to speed up training of ANNs \citep{LeCun2012}. The values of the mean and standard deviation are derived during training and are not changed when the model is tested on the validation data.

It is also interesting to compare using ANNs in this way against using ANNs that are trained to replace dynamical models or components of them, as done by \citet{Dueben2018}. Therefore multilayer perceptron ANNs were also trained to predict the full tendency of the Truth system $\mathrm{d}X_{k}/\mathrm{d}t$, given the same inputs as the ANN error-correctors. 
Again, these predict the tendency for one $X$ variable at a time. In an Earth system model, individual parameterisation schemes could also be replaced by ANNs. However, the Lorenz '96 system lacks the complexity to do an interesting experiment where something closely analogous to replacing a parameterisation scheme is carried out, given the simplicity of $U(X)$.

ANNs with different arrangements of neurons were tested, with one or more hidden layers (the ``depth'') and with an equal number of neurons in each hidden layer (the ``width''). As a shorthand notation, an ANN with depth $D$ and width $W$ will be referred to as ``d$D$w$W$'' (e.g. d2w32 refers to an ANN with depth 2 and width 32). The ANNs all use a linear output activation function and rectified linear unit activation functions on the hidden layers, which were found to work more robustly than hyperbolic tangent functions for the case of training ANNs to predict the full system tendencies. For this case, the outputs of ANNs with hyperbolic tangent activation functions were found to be prone to saturating, so that the largest tendencies could not be simulated. This may have happened because the magnitude of the output is limited to be the sum of the magnitudes of the weights connecting the final hidden layer to the output, which were not made large enough in training. This suggests that for predicting values that can take any size, using activation functions whose output values are not typically limited in magnitude is more likely to give good results.

\subsubsection{Training ANNs}
Results are presented for models using ANNs trained on 1000MTUs of truth model data, using the tendency over every 0.005MTU interval, in order to test their potential skill when data availability is not a limitation. Using all 3000MTUs of the training data was not found to increase the skill of ANNs substantially when tested on a few chosen ANN structures. It is also shown in section~\ref{sec:res:ten_errs} that the performance of the ANNs is similar at predicting tendencies in the training and validation truth datasets, indicating that the ANNs are not substantially overfitting the training data, so increasing the amount of training data would not be expected to improve the ANNs' performances much. 

ANNs are trained to minimise the sum of the squared prediction error and an $L_{2}$ regularisation term for the weights with coefficient $10^{-4}$. This was done using stochastic gradient descent with the Adam algorithm \citep{Kingma2014}. Minibatches of size 200 sets of input and output were used together with a learning rate of 0.001. Training stopped when the squared prediction error failed to decrease by at least $10^{-4}$ twice consecutively after iterating over the whole training dataset.

\subsection{Results} \label{sec:res}
Diagnostics comparing simulations from the Truth, No-ANN model and models using ANNs are shown below. All results are very robust to sampling variability, as determined by checking that they are very similar when only half of the data is used, except for the difference in the mean biases (section~\ref{sec:res:fc_clim_skill}) for which the use of ANNs was not found to give a statistically significant difference in most cases.

\subsubsection{One-timestep tendency forecast errors} \label{sec:res:ten_errs}
Figure~\ref{fig:ten_err} shows the root mean square error (RMSE) of tendency predictions over a single coarse time step (0.005MTUs) for coarse-resolution models with error-correcting ANNs . Results are shown for models with ANNs with depths up to three and widths that are integer powers of 2 between 2 and 64 (width-1 ANNs did not generally perform well, as would be expected). The RMSE is calculated for 10,000 randomly chosen time steps in each of the training and validation datasets, using the same time steps for each ANN. The error is reduced compared to that for the No-ANN model for every ANN structure, showing that even very simple ANNs (e.g. with two neurons in a single layer) can improve the skill of predicting the tendencies. The errors do generally decrease as the ANN width and depth each increase, indicating that the optimal function relating the coarse-resolution model's tendency errors to the $X$ variables may be quite complex. The maximum error reduction on the validation dataset is 42\% for the largest (d3w64) ANN, showing that ANNs can greatly reduce the error. The RMSEs on the validation dataset are not more than 3\% above those on the training dataset, indicating that no substantial overfitting is occurring. The errors decrease as ANN size increases, and it seems likely that futher error reductions are possible. The aim here is to explore how it might be made easier to get some improvement using ANNs, rather than to find the optimum performance, so results for larger ANNs and varied training hyperparameters are not shown.

For comparison, figure~\ref{fig:ten_err_full} shows RMSEs of tendency errors of coarse-resolution models using ANNs to predict the full tendency. Errors are generally higher than for the models using error-correcting ANNs for a given width and depth, and are only better than the No-ANN model once the ANNs become sufficiently large (with width at least 32 or width at least 16 with a depth of two or more). This illustrates how more complex ANNs are generally required to replace the model components rather than just to correct their errors, making it harder to achieve better performance using this approach. More parameters are also needed, increasing the risk of overfitting the data. Note that it could still be the case that the optimum attainable performance, using ANNs larger than those tested here, is better than in models with error-correcting ANNs.

Improvements upon the No-ANN model can also be obtained with much smaller amounts of training data using the error-correcting approach. Models with ANN correctors trained on just 2MTUs of training data predicted tendencies for the validation dataset with root mean square errors (RMSEs) that were robustly less than those predicted from the No-ANN model in most cases (not shown). Pure ANN models require at least about 20MTUs to achieve this. For comparison with the typical time scales of the true system, the autocorrelation of the $X$ variables falls to about 0.05 after a lag of 0.4MTUs. 

In order to determine if there are any situations in which the errors of tendencies predicted by the models using error-correcting ANNs are large, figure~\ref{fig:ten_pred} shows scatter plots of predicted tendencies and their errors versus the true tendencies. The sets of tendencies shown are from the No-ANN model and two example coarse-resolution models with error-correcting ANNs (with d1w16 and d2w32 structures, the latter performing particularly well at improving short-term forecast and climate skill scores [section~\ref{sec:res:fc_clim_skill}]). 100,000 scatter points are shown for tendencies predicted in each of the training and validation datasets. 

The models with error-correcting ANNs predict tendencies that are close to the true tendencies in both the training and validation datasets, including for extreme positive and negative tendencies. The predicted tendencies are generally closer to the true tendencies than those made by the No-ANN model throughout the whole range of true tendency values, including for extreme cases, although the No-ANN model also does not make any particularly large errors. This is evidence that the ANNs have learnt how to actually improve the representation of the dynamics, so that they can improve most predictions and not degrade predictions of extreme values in the validation dataset even when there are few examples of the latter in the training data. This is generally the case for all of the different ANN structures, even for the smallest ANN that was tested (d1w2; not shown). It is important to show that ANNs do not simply fit the training data and perform poorly at extrapolating to make predictions for rare, extreme situations, since it is essential in Earth system modelling applications that models' performance does not severely degrade in these cases.

\subsubsection{Forecast and climate simulation skill} \label{sec:res:fc_clim_skill}
\begin{figure}
 \centering
 \includegraphics[width=\textwidth]{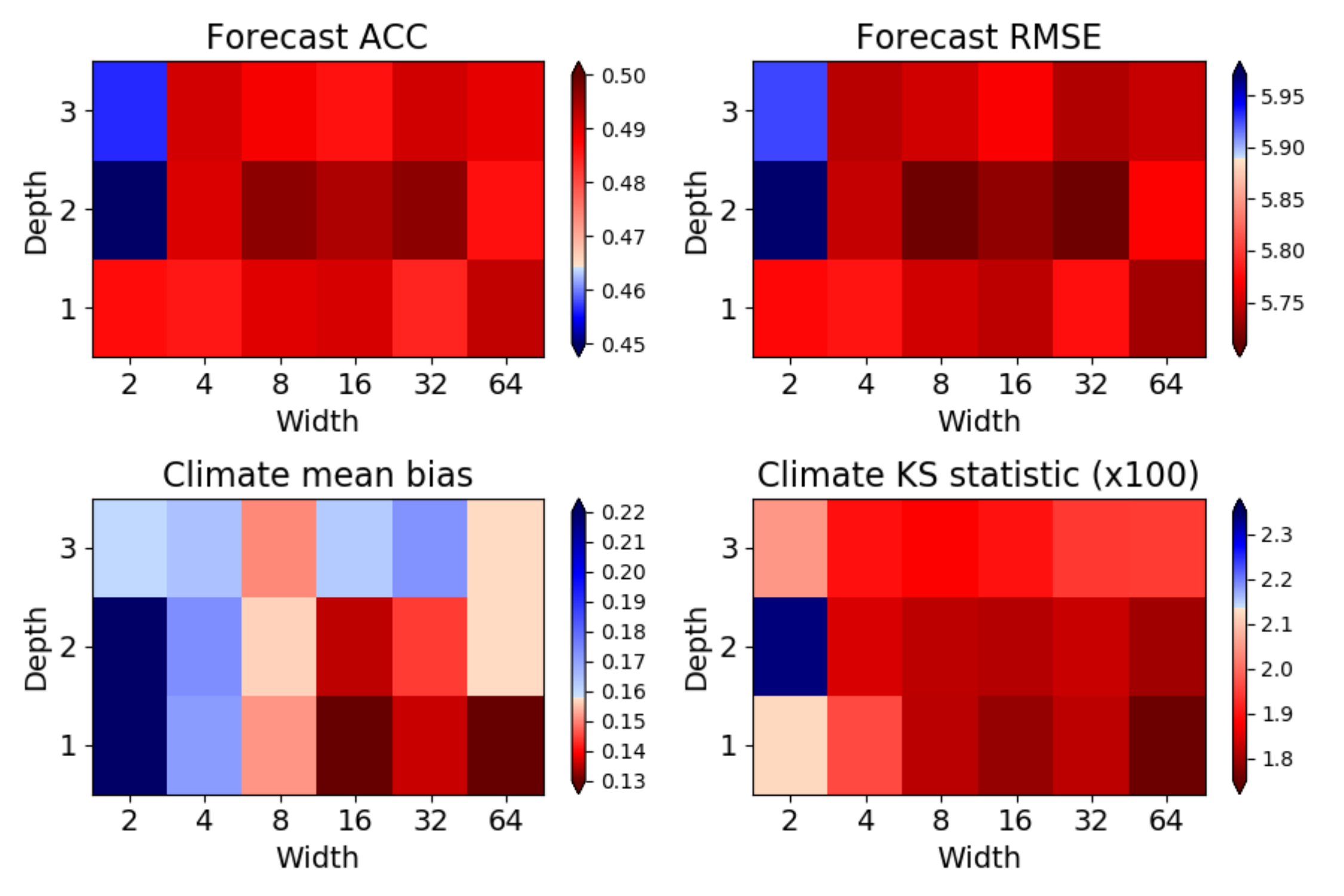}
 \caption{Forecast skill and climate simulation diagnostics for models with error-correcting ANNs with different widths and depths, evaluated using the validation dataset as a reference: the forecast anomaly correlation coefficient (ACC; top left) and root mean square error (RMSE; top right), both at lead time 1MTU, and the climate time-mean bias (bottom left) and Kolmogorov-Smirnov (KS) statistic (bottom right) of the $X$ variables. Red indicates better performance relative to the No-ANN model and blue worse performance. The models with ANNs generally have a better ACC, RMSE and KS statistic than the No-ANN model, but most differences in the mean bias were not found to be statistically significant (see text).}
 \label{fig:fc_clim_diags}
 \end{figure}

Metrics of forecast skill and the quality of the simulated climate of the $X$ variables are shown in figure~\ref{fig:fc_clim_diags} for coarse-resolution models with error-correcting ANNs of different depths and widths, evaluated using the validation dataset only (this is the case for all model quality metrics shown from now on). 

Forecast diagnostics were computed from 10-member ensembles of simulations initialised from each of 3000 states of the $X$-variables sampled from the Truth validation run, each separated by 1MTU, giving effectively-independent initial conditions. To form the initial conditions for each ensemble member for each Truth initial condition, random perturbations were sampled for each $X$ variable independently (noting that correlations between $X$ variables in the Truth system are small). Firstly, a sample ($\mu$) was taken from a Gaussian distribution with a mean of zero and a standard deviation of 0.05. Then ten samples were taken from a Gaussian distribution with a mean $\mu$ and a standard deviation of 0.05 and added to the Truth state. This ensured that the population standard deviation of the initial conditions equalled the standard deviation of the differences between their means and the Truth states, as would be expected if the perturbations came from a well-calibrated error distribution in the estimate of the initial state in a forecasting system.

The forecast anomaly correlation coefficient (ACC) and RMSE at lead time 1MTU are better than in the No-ANN model for all models with ANNs except those with width 2 and depth 2 or 3 (figure~\ref{fig:fc_clim_diags}, top; squares are shaded red where the metric is better than that for the No-ANN model). (Forecasts at a lead time of 1MTU are roughly analogous to a ``medium-range'' forecast of the Earth's atmosphere, given the autocorrelation time scale of the system). Therefore in most cases the improvement in representing the single timestep tendencies (section~\ref{sec:res:ten_errs}) has brought about an improvement of longer range forecast skill relative to the No-ANN model. The improvement seems to be quite modest, however, raising the ACC from about 0.46 to 0.49 and decreasing the RMSE from 5.89 to 5.73 at best. However, note that the ACC for the Truth model initialised with the same initial conditions perturbations is only 0.52 and its RMSE is 5.59. This is the maximum potential skill. Therefore the best improvements in the ACC and RMSE are slightly over 60\% of the difference between the maximum possible skill and that of the No-ANN model. For the median case, they are 49\% and 48\% of the difference respectively. This suggests that in a case where forecast skill were much lower than the maximum possible skill than it is here, the absolute skill improvements gained by using ANNs could be much more substantial. (Indeed, ANNs that are trained to simulate the full tendency $\mathrm{d}X_{k}/\mathrm{d}t$ can have skill similar to the models with error-correcting ANNs tested here [not shown]. This suggests than ANNs could learn to correct the errors of a No-ANN model that were degraded to have a much lower skill level, so that the gap between the No-ANN model and the models with ANNs were much larger, though this is not tested here.)

The biases of the time-mean of the $X$ variables diagnosed from 3000MTU climate runs are shown in the bottom-left panel of figure~\ref{fig:fc_clim_diags}. The diagnosed biases are mostly similar to those of the No-ANN model, except those for the models with d1w2 and d2w2 ANNs, which have much larger biases. This is the one diagnostic for which sampling variability is substantial. The biases are not statistically significantly different from that of the No-ANN model at the 95\% level, except in the cases of the models with d1w2 and d2w2 ANNs. Therefore it is difficult to be confident about how many of the models with error-correcting ANNs have smaller mean biases without using much longer climate runs, but it seems clear that the changes in the bias are quite small overall. (The statistical significance was calculated according to a Monte Carlo permutation test \citep{Efron1994}. Each time series was divided into blocks of length 100MTU, which is much larger than the autocorrelation time scale of the data.  For each model with an ANN, surrogate time series of length 3000MTU were created by selecting blocks randomly without replacement from the simulation by this model and the simulation by the No-ANN model. The probability of the absolute difference between the means of two of these time series being smaller than that between the actual time series was calculated to quantify the statistical significance.)  

In order to evaluate improvements in the shape as well as the mean of the simulated climatological distribution of $X$ values, the two-sample Kolmogorov-Smirnov (KS) statistic was calculated between the simulated distribution and the distribution in the truth model validation run. This is simply the maximum difference between the cumulative density functions of the two distributions as a function of $X$. The KS statistic is improved in all but the d2w2 case, by up to $\sim$15\% (figure~\ref{fig:fc_clim_diags}, bottom right). Part of the reason that this happens even though the bias in the mean of the distribution is not always improved is that the variance of the $X$ values is increased relative to that in the No-ANN model (not shown), bringing the $X$-distribution closer to that of the truth model by this measure, except in the d2w2 case. 

Formal statistical significance tests were not carried out for diagnostics other than the mean bias because it seems very unlikely to get the result that models with all but the smallest ANNs seem to have improved diagnostics (figure~\ref{fig:fc_clim_diags}) if it were not the case that ANNs were truly producing improvements in most cases. Detailed consideration of the sampling uncertainty would be required to assess the relative skill for different ANN structures, but it is not the aim here to do this.

Altogether this indicates that the use of error-correcting ANNs in this system is able to robustly give improvements in forecast skill and the shape of the climate distribution relative to that of the No-ANN model. However, comparing figure~\ref{fig:fc_clim_diags} with figure~\ref{fig:ten_err} shows that improving the error of the predicted tendency does not guarantee that the quality of longer simulations will also improve. The improvements in the climate diagnostics are also smaller than might be anticipated, given the large reduction in the tendency errors that was shown in section~\ref{sec:res:ten_errs}.

\begin{figure}
 \centering
 \includegraphics[width=\textwidth]{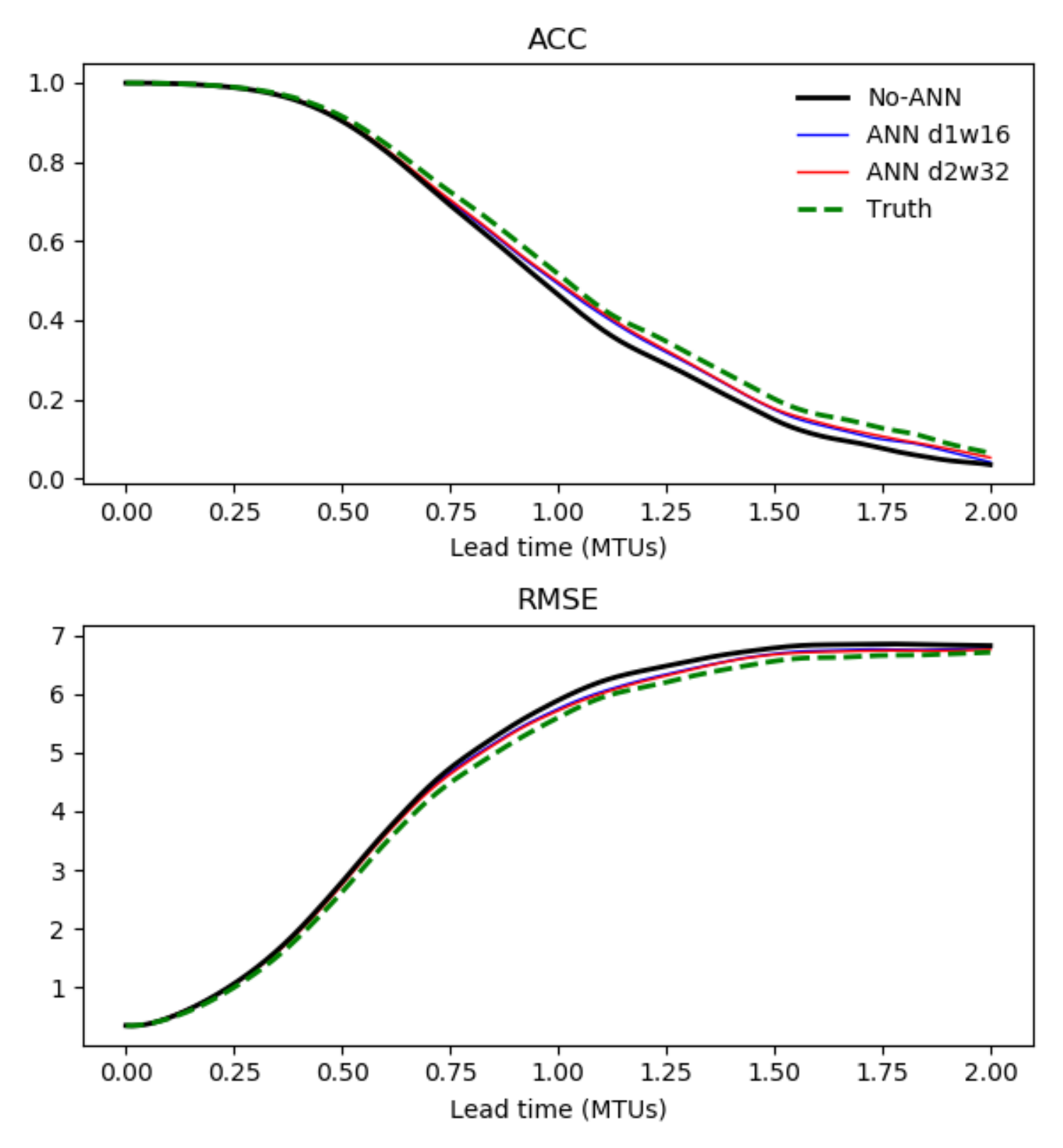} 
 \caption{Forecast skill as a function of lead time evaluated on the validation dataset: the anomaly correlation coefficient (top) and root mean square error (bottom). Results are shown for the No-ANN model, coarse-resolution models with with depth-1 width-16 and depth-2 width-32 error-correcting ANNs and the Truth model, which shows the maximum potential skill with the given initial condition perturbations. The models with ANNs give modestly higher skill at lead times greater than about 0.75MTUs, approximately halving the difference relative to the Truth model's skill.}
 \label{fig:fc_diags_good_nns}
 \end{figure}

Figure~\ref{fig:fc_diags_good_nns} shows the forecast ACC and RMSE as a function of lead time for the No-ANN model and the models with the d1w16 and d2w32 error-correcting ANNs, which are the same models that were used for figure~\ref{fig:ten_pred}. The forecast skill is very similar for the different models up to a lead time of about 0.75MTUs, after which the models with error-correcting ANNs begin to have higher skill than the No-ANN model. After a lead time of $\sim$1MTU, their skill is approximately half way between that of the No-ANN model and the Truth model using the same initial condition perturbations. The maximum skill differences between the models with the error-correcting ANNs and the No-ANN model are about 0.04 in the ACC and 0.2 in the RMSE.

\begin{figure}
 \centering
 \includegraphics[width=\textwidth]{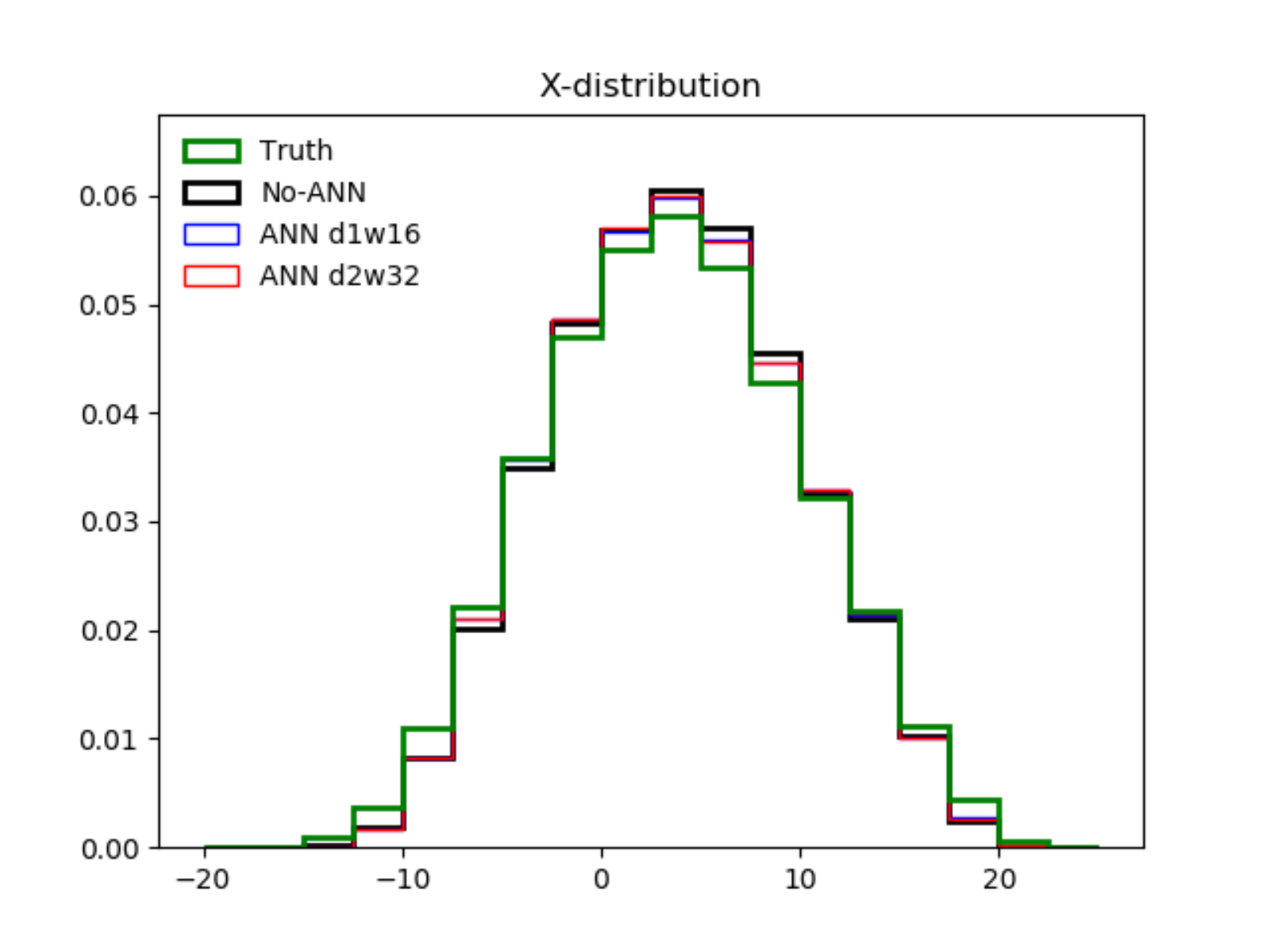}
 \caption{Frequency distribution of $X$ values in long simulations in the validation dataset ("Truth") and in several coarse-resolution models: the No-ANN model and those with error-correcting ANNs with depth-1 and width-16 and with depth-2 and width-32. The simulations by the models with ANNs have a smaller excess of frequencies of values near the centre of the distribution than the No-ANN model, but all have too low a frequency of extreme values.}
 \label{fig:clim_hist}
 \end{figure}

To understand better how the improvements in climate statistics shown in figure~\ref{fig:fc_clim_diags} are manifested in the frequency distribution of the $X$ variables, figure~\ref{fig:clim_hist} shows their distribution in the Truth validation dataset, in the No-ANN model and in the previously discussed models with the error-correcting ANNs. The simulations produced by the latter have smaller frequencies near the centre of the distribution, so that the bias here is smaller, with the frequencies at moderate negative values between about $-7.5$ and $-5$ beneficially increased. All of the coarse-resolution models have too low frequencies of large positive and negative $X$-values, however. This may indicate that it is not possible to simulate the correct frequencies of these extremes without explicitly representing the $Y$ variables, though it is also possible that it could be improved by applying better machine-learning approaches or including stochasticity in the coarse-resolution models \citep{Arnold2013}.

\begin{figure}
 \centering
 \includegraphics[width=\textwidth]{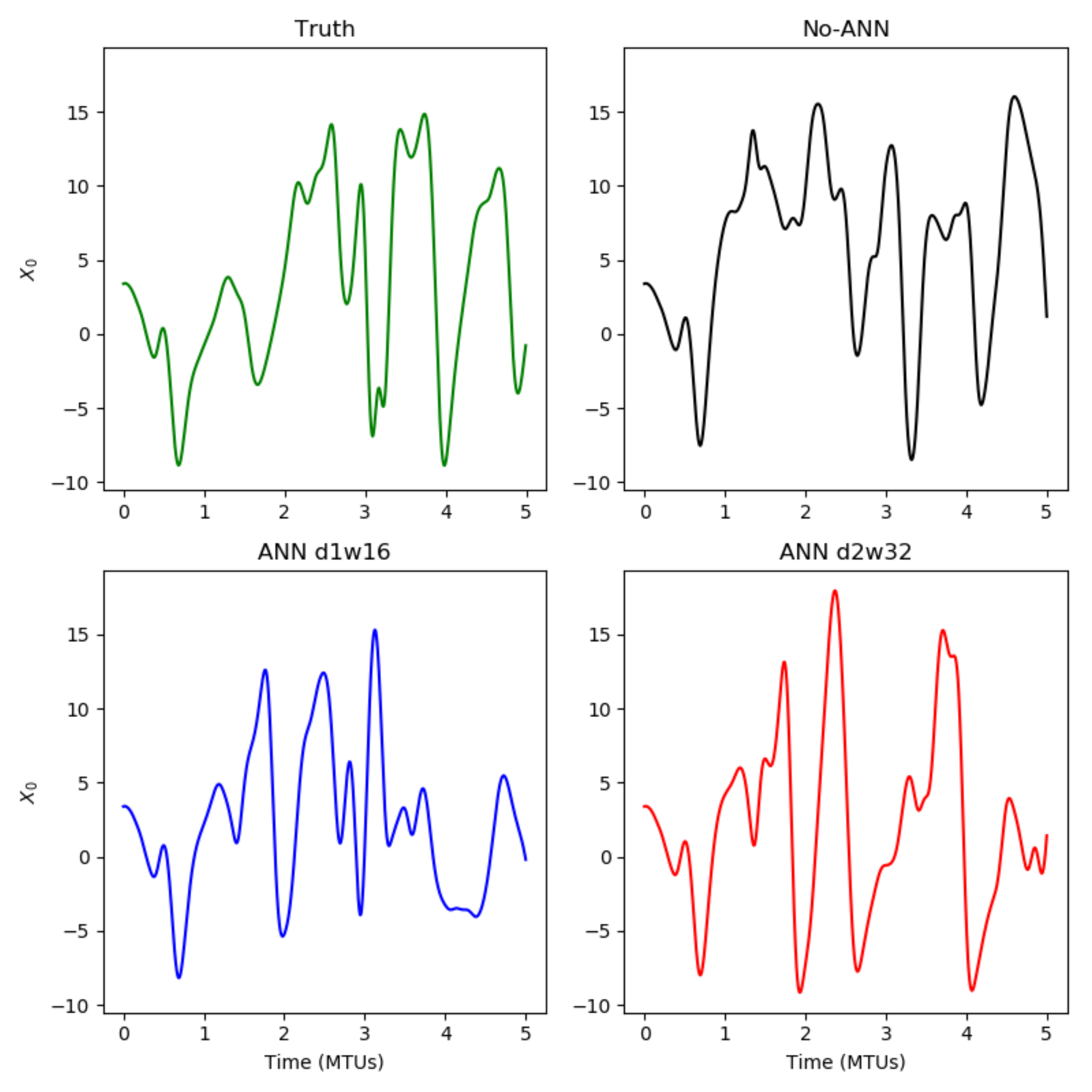} 
 \caption{Time series of the first $X$ variable in the validation run (top left) and in different coarse-resolution models initialised with the state of the validation run at time zero: the No-ANN model (top right) and models with error-correcting ANNs with depth-1 width-16 (bottom row, left) and depth-2 and width-32 (bottom row, right). The models with ANNs capture the initial evolution of the true system, and then beyond the predictability limit they exhibit similar variability to the true system.}
 \label{fig:ts}
 \end{figure}

On top of considering statistical summary measures of simulation skill, it is also important to verify that the temporal evolution of the system state is realistically simulated in the models with ANNs. Figure~\ref{fig:ts} shows a time series of the first $X$ variable of length 5MTUs at the start of the validation dataset ("Truth"; figure~\ref{fig:ts}, top left), in the No-ANN model (top right) and in the models with the d1w16 and d2w32 error-correcting ANNs (bottom). All time series begin with the initial condition of the validation data at time zero, so that the models capture the features of the initial evolution up to about time 1MTU, and then the model simulations diverge, likely primarily due to chaotic variability. After this point, the coarse-resolution models produce variability that appears qualitatively similar to that in the Truth system.

To quantify the impact of using $X_{k+2}$ as an input to the ANNs, which is not used as an input to the No-ANN model, results for error-correcting ANNs using the same inputs as the No-ANN model were analysed. These were also found to robustly improve the metrics discussed above (again, with the exception of the mean climate bias). The improvements in short-range prediction errors relative to the No-ANN model were smaller than found above, by a median of 18\% for one-timestep tendency errors across ANNs with the same structures as those considered above and by about 40\% for the ACC and RMSE at a lead time of 1MTU. The improvements in the climate KS statistic tended to be slightly greater, however, by a median of 5\%---this is perhaps because optimising short-range skill does not always optimise long-range skill. Overall, this illustrates that using inputs that are not presently incorporated in all Earth system model components could be a substantial source of skill added by machine learning algorithms (for example, data in horizontally-adjacent grid columns that is not typically used in subgrid parameterisation schemes).

\section{Discussion} \label{sec:disc}

\subsection{Additional considerations for Earth system modelling} \label{sec:disc:esm}
The approach used here of training ANNs to reduce single-timestep tendency errors could not be applied exactly analogously to learn to better represent the dynamics of the Earth system because observations at a given location are typically spaced six hours or more apart, and state-of-the-art dynamical Earth system models use time steps that are much shorter. Maintaining a short time step is desirable so that the model equations can better approximate the true equations, which are continuous in time. It may also be necessary to ensure numerical stability. Therefore an approach is required that could update the learning algorithm's parameters based on what would improve forecast skill over multiple time steps. \citet{Brenowitz2018} achieve this when emulating convection-resolving simulations in a single column model by optimising a cost function that takes into account errors in a prediction over multiple time steps---this is in order to make their system stable, but it may also help to improve longer range prediction skill. In a free-running system, the impact of parameter perturbations on output from previous time steps would need to be taken into account, on top of their impact in the time step corresponding to the observation. The ``backpropagation through time'' algorithm \citep{Werbos1990} that is used in recurrent ANNs \citep{Funahashi1993} could be used. 
In a model with multiple grid points, if the Earth system learning algorithm is ``local'', then the effect of varying the algorithm's parameters on predictions at nearby grid points probably needs to be taken into account as well as the effect on the predictions through multiple time steps---the backpropagation needs to be done ``backwards through time and sideways through space''. This is because tendencies at a given grid point depend on the system state at nearby grid points, and so prediction errors at those points at earlier time steps need to be accounted for. (It seems desirable for the algorithm to take inputs only from local grid points in order to be easier to implement in parallel computing environments \citep{Dueben2018} and to respect symmetry of the physical equations with respect to spatial translation.) For error-correcting algorithms, this approach requires the tangent linear approximation of the remainder of the model, which is related to the adjoint models that are often used in data assimilation \citep{Errico1997}, and have been developed for some models of Earth system components \citep[e.g.][]{Janiskova2013,Lea2015}. 

The data used for training algorithms also needs to be considered. \citet{Dueben2018} suggest using reanalysis data. Although reanalysis data is imperfect, it is likely to have smaller climate biases than existing dynamical models, enabling the algorithms to yield performance improvements. A possible next step would be to recalculate the reanalyis using the improved model, combining information from this model and observations to get a yet better estimate of climate statistics. This could then be used to train better algorithms, and so on, yielding further upward steps in performance, as well as an optimal estimate of past weather given our observations.

\subsection{Application to problems beyond increasing prediction skill with a stationary climate}
Error-correcting algorithms in dynamical models may be useful for addressing problems besides improving simulation skill. For example, if they do a good job at correcting large model errors, then it may be possible to understand from them how model components like conventional parameterisation schemes can be improved, making use of advances in interpreting the workings of algorithms like ANNs \citep[e.g.][]{Ribeiro2016}. They could also help to constrain stochastic parameterisations \citep{Palmer2001,Watson2015a} by placing an upper bound on the size of the component of the tendencies that is not predictable given the variables on the coarse grid, an irreducible error for a given model resolution, which can be modelled stochastically. Generative-adversarial algorithms \citep{Goodfellow2014} could also find better ways to model the effects of unresolved flow stochastically [\textit{Xie et al.}, 2018; \textit{Gagne~II, DJ et al.}, ``Machine Learning for Stochastic Parameterization: Generative Adversarial Networks in the Lorenz '96 Model'', in preparation] \nocite{Xie2018}.

The ability to vary the complexity of algorithms like ANNs in a systematic way to create a model ensemble also allows for testing of the seamless prediction paradigm---the idea that models that have better short-range prediction skill also have better long-range skill, which would mean that metrics of weather forecast skill would be informative about models' abilities to simulate the climate response to anthropogenic forcing \citep{Palmer2008,Matsueda2016}. Alternative methods of creating an ensemble of models such as by perturbing model parameters may generally struggle to give any skill improvements, so it cannot be seen if climate simulation skill improves as short-range prediction skill gets better. In the Lorenz '96 system studied here, correlations between the single-tendency prediction error in the validation dataset (figure~\ref{fig:ten_err}, right) and the forecast and climate skill diagnostics shown in figure~\ref{fig:fc_clim_diags} have magnitudes between 0.63 and 0.70. Correlations between the forecast RMSE at lead time 1MTU (figure~\ref{fig:fc_clim_diags}, top right) and the climate mean and KS statistic (figure~\ref{fig:fc_clim_diags}, bottom panels) are 0.57 and 0.81 respectively. This quantifies the relationship between short-range and long-range skill in this system when using error-correcting ANNs, showing that improvements in predictions at shorter lead times do indeed tend to be associated with improvements in long-range predictions. However, as noted earlier, the correspondence is not perfect, and the improvements made to long-term climate diagnostics by using ANNs are considerably smaller than what might be expected given the improvements made to single-timestep tendency predictions. Therefore the seamless prediction paradigm does not apply fully. It would be very interesting to see how well it applies in Earth system models, given the correspondence that has been identified between biases in short-range forecasts and simulated climate \citep{Ma2013,Sexton2019}.

Another interesting question is whether using statistical learning algorithms within Earth system models could help to give more accurate simulations of the impacts of anthropogenic climate change. This is challenging because this requires making predictions about conditions that are dissimilar from those we have observed, so that a good representation of the underlying dynamics of the system is necessary. \citet{OGorman2018} found that their emulation of a convection parameterisation could not reproduce the effect of climate change well when it was trained only in a stationary ``control'' climate. However, statistical approaches such as optimal fingerprinting are well-established in work on detection and attribution of climate change and can be used to estimate the extent to which a given model is over- or under-estimating the response to a particular forcing \citep{Bindoff2013}. The climate change signal in individual weather events also appears clearer when dynamical variability is controlled for, which has been done previously using weather analogues \citep[e.g.][]{Yiou2007,Cattiaux2010}. This suggests that there is scope for learning the effects of anthropogenic emissions more precisely within a model that can also accurately take into account all of the other influences on individual weather events. Even if such a model would not be trusted for projecting the impacts of large climatic changes without people being able to understand the calculations behind its predictions, it may still be useful for problems such as the attribution of observed extreme weather events \citep{Allen2003,NASEM2016}, for which extrapolation beyond observed conditions is not so much of a concern.

\section{Conclusions}
It has been shown that artificial neural networks can learn to correct errors of a coarse-resolution model of a chaotic dynamical system (the Lorenz '96 system), resulting in stable simulations that have both improved skill in initialised forecasts and better long-term climate statistics. Improvements are found for a wide range of ANN structures, showing that they are quite robust.  

\begin{figure}
 \centering
 \includegraphics[width=\textwidth]{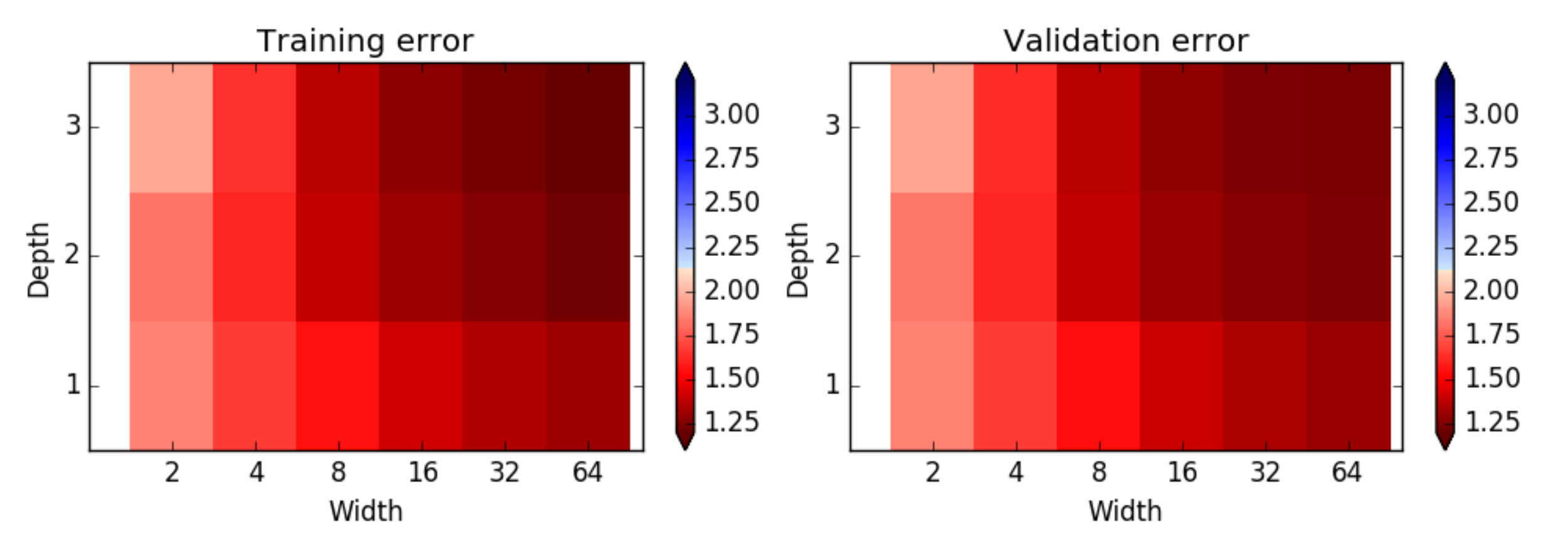}
 \caption{The root mean square error of the tendency predictions over one coarse time step (0.005MTUs) of coarse-resolution models with error-correcting ANNs with different widths and depths evaluated on the training data (left) and validation data (right). Red indicates better performance relative to the No-ANN model and blue worse performance (with the value for the No-ANN model being the value at which the colour changes). The ANNs give robust outperformance of the No-ANN model.}
 \label{fig:ten_err}
 \end{figure}

\begin{figure}
 \centering
 \includegraphics[width=\textwidth]{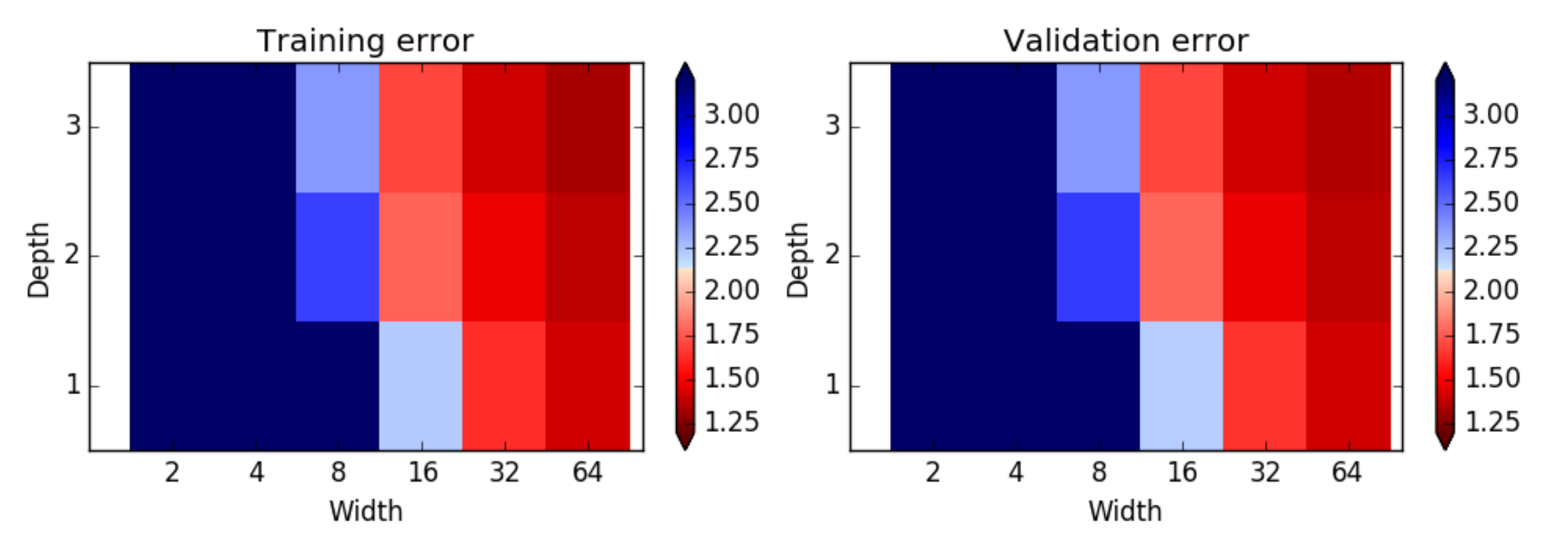}
\caption{RMSEs of tendency errors, as in figure~\ref{fig:ten_err}, but for ANNs predicting the full $X$ tendencies (section~\ref{sec:L96exp:ANNmods}). Note that the colour scale is saturated for the smallest ANNs. More complex ANNs are required to outperform the No-ANN model than for models using error-correcting ANNs.}
\label{fig:ten_err_full}
 \end{figure}

\begin{figure}
 \centering
 \includegraphics[width=\textwidth]{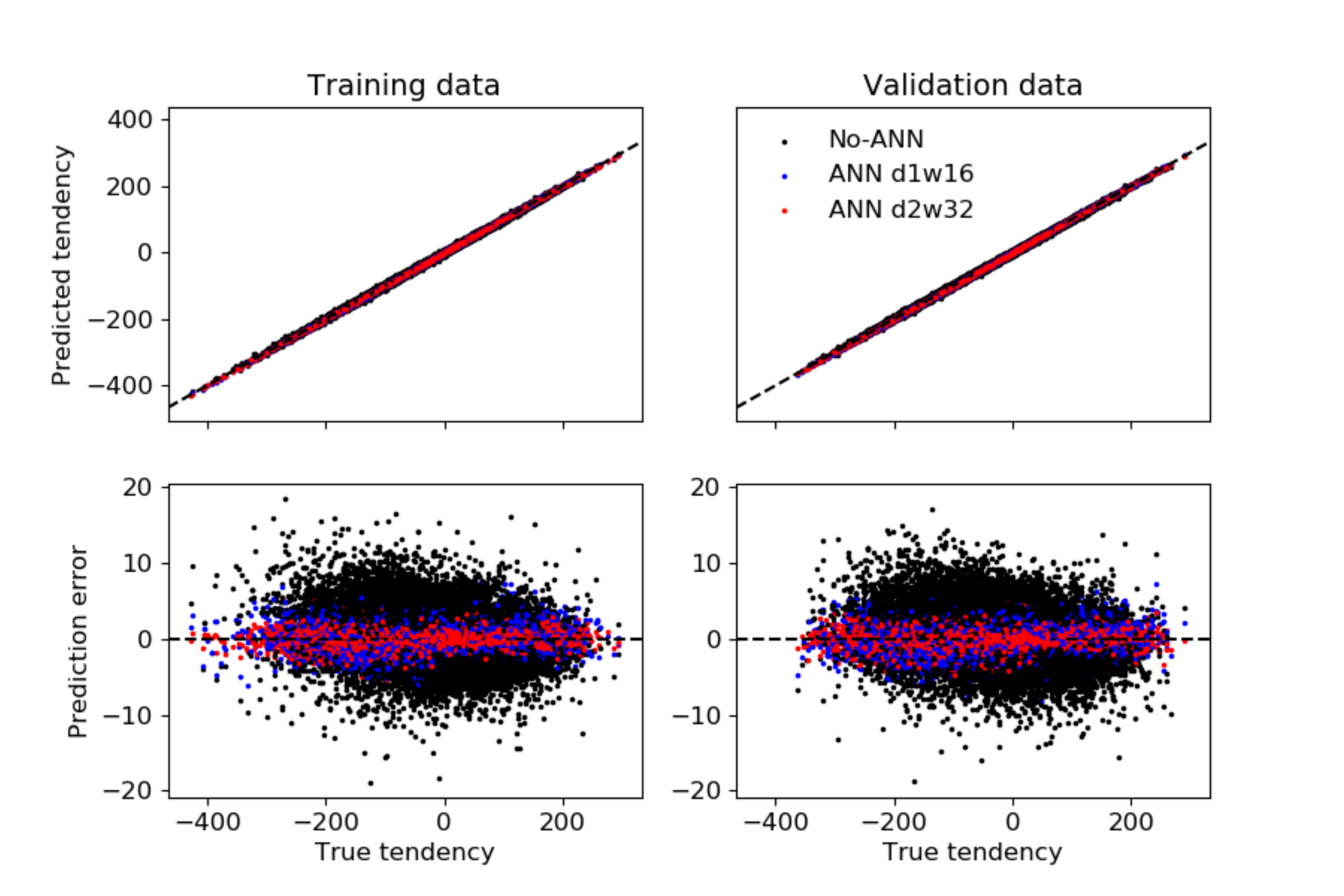}
 \caption{The top panels show the tendencies predicted by several coarse-resolution models plotted against the true tendencies in the training dataset (top left) and validation dataset (top right). The models are the No-ANN model and the models with depth-1 width-16 and depth-2 width-32 error-correcting ANNs. The lower panels show the prediction errors plotted against the true tendencies. The dashed lines indicate the values for perfect predictions. The tendency predictions for the models using ANNs are mostly closer to the truth than the predictions from the No-ANN model, including for rare extreme values in the validation dataset, which is evidence that the ANNs have learnt to improve the representation of the dynamics.}
 \label{fig:ten_pred}
 \end{figure}

The ANNs used here could reduce errors in single time step predictions by up to about 40\%, and it seems that the errors could be reduced yet further if the ANNs were increased in size (figure~\ref{fig:ten_err}), though it is not the aim here to find the best possible performance. Errors of predicted single-timestep tendencies become gradually smaller as the ANN complexity increases (figure~\ref{fig:ten_err}), and there does not appear to be a substantial problem due to the training getting stuck in poor local minima. The models with ANNs also give good predictions of extreme tendencies that were not seen in the model training stage (figure~\ref{fig:ten_pred}). 

In initialised ``medium-range'' forecasts, the improvement in the absolute anomaly correlation coefficient and root mean squared error was only a few percent at lead times of $\sim$1MTU. However, this was $\sim$50\% of the maximum possible improvement in the median case, determined by comparison with forecasts made by the Truth model with the same initial condition perturbations. The improvement of climate statistics was modest, with improvements up to $\sim$15\% in the climate Kolmogorov-Smirnov statistic and no discernible improvement in the time-mean state. 
This may be because the model without an ANN was already actually quite skilful at predicting the Truth system's behaviour---for example, figure~\ref{fig:ten_pred} shows that its predicted tendencies are always quite close to the true tendencies. For models of Earth's atmosphere, coarse-graining studies find much worse agreement between tendencies predicted by models and estimated true tendencies \citep[e.g.][]{Shutts2007,Shutts2014}, suggesting that there may be much more room for improvement using machine learning. However, as discussed in section~\ref{sec:disc:esm}, getting large benefits in longer range skill may require training algorithms to target improvements on time scales longer than single time steps.

These results support the idea that ANNs (or other machine learning algorithms) could help to reduce errors in dynamical Earth system simulations by learning a better representation of the physical equations from observations or from more realistic models that are too expensive to use generally in weather and climate prediction \citep{Dueben2018}. However, it can be far easier to use ANNs to correct the output of an existing model than to train ANNs to simulate the entire system, because far smaller ANNs can be used and much less data is required for training the ANNs (it was also shown by \citet{Jia2018} that error-correcting ANNs require less data, in the context of simulating lake temperatures). Also, Earth system models typically relate dozens of inputs and outputs at every grid point, but an error-correcting system can produce performance improvements whilst only considering a subset of the models' inputs and outputs, meaning it is possible to begin demonstrating improvements without reproducing the complexity of the full model. This is valuable because the more complex the ANN that is required, the harder it is generally to find a training method that produces good results. This method also utilises the physical understanding embedded in the existing parameterisation schemes, and the error-corrections should only become large in situations when the schemes do not perform well, reducing concerns about their reliability. This makes this approach more appropriate for use in a research program to investigate the potential for ANNs to reduce model errors and to begin producing operational improvements. The next step is to test whether the method works as well in models of components of the Earth system.

The main drawback of this approach compared to training an algorithm to simulate the full system is that the computational cost of the model cannot be reduced. Using algorithms like ANNs to learn to represent the full system's dynamics may therefore be the approach adopted in the long run, but developing systems to learn to correct model errors will give invaluable insights about how to achieve this in the medium term, and help to demonstrate whether attempting to learn a better representation of the full dynamics from observations or expensive models is likely to give a substantial improvement in forecast skill. (There is also nothing to preclude an error-correcting algorithm being used in conjunction with emulators of an existing model's parameterisation schemes or high-resolution simulations that do reduce the computational cost \citep[e.g.][]{Chevallier1998,Krasnopolsky2010a,Brenowitz2018,Gentine2018,Rasp2018a,OGorman2018}.) Models of the Lorenz '96 system using ANNs to predict the full tendency were found to achieve similar performance to the models with error-correcting ANNs, just requiring larger ANNs to do so (not shown). Therefore there is nothing in the results presented here to preclude using ANNs in place of physically-derived models eventually. The two methods can be used in a complementary way in a research programme.

\section*{Acknowledgments}
I thank Peter Dueben and members of Tim Palmer's research group, particularly Matthew Chantry and Jan Ackmann, for stimulating discussions about this work, and also Myles Allen, Tim Woollings and Tim Palmer for supervisory support. I also thank two anonymous reviewers for their constructive comments. I received funding from European Research Council grant 291406 and Natural Environment Research Council grant NE/P002099/1. No external data sources are required to reproduce the results presented in the manuscript. A Jupyter notebook for training and evaluating the models with ANNs can be found at https://github.com/PAGWatson/Lorenz96\_and\_neural\_networks.

\clearpage

\clearpage


\end{document}